\documentstyle[11pt,aaspp4]{article}

\begin{document}
\title{A NICMOS Survey of Early-Type Galaxy Centers$^1$: \\
The Relation Between Core Properties, Gas and Dust Content and Environment}

\author{
A.~C.\ Quillen,\altaffilmark{2}$^,$\altaffilmark{4}
Gary A.\ Bower\altaffilmark{3} \&
M.\ Stritzinger\altaffilmark{2}
}
\altaffiltext{1}{Based on observations with the NASA/ESA {\it Hubble Space Telescope}
obtained at the Space Telescope Science Institute which is operated by the 
Association of University for Research in Astronomy, Inc. (AURA), under 
NASA contract NAS5-26555.}
\altaffiltext{2}{The University of Arizona, Steward Observatory, Tucson, AZ 85721} 
\altaffiltext{3}{Kitt Peak National Observatory, National Optical
Astronomy Observatories, P.O. Box 26732, Tucson, AZ 85726; gbower@noao.edu.}
\altaffiltext{4}{aquillen@as.arizona.edu}

\begin{abstract}
We present a NICMOS 1.6 micron imaging isophotal study of 
27 early-type galaxies.  Our selected sample is not as biased against dusty
galaxies as the visible WFPC samples observed with HST.
Core galaxies have reduced ellipticity and boxiness
near and within their core or break radius.  
This supports a core formation mechanism which mixes or scatters 
stars such as scattering caused by a binary black hole.
We find the same trends between central surface brightness
and luminosities as the WPFC studies.
We find no correlation between core properties 
and dust mass or X-ray luminosity, suggesting
that processes determining the current gas content 
(e.g., such as minor mergers and cooling flows)
are unrelated to processes occurring during core formation.  
Core galaxies exist in a variety of environments ranging
from poor groups to large clusters.
A combined sample suggests that galaxy groups may harbor more luminous 
power law galaxies than clusters such as Virgo and Fornax.
\end{abstract}


\section{Introduction}
Optical imaging surveys carried out with The Hubble Space Telescope (HST)
find that almost all early-type galaxies harbor dust (\cite{vdb94}).
Even though projection effects cause a bias against detecting
dust features in most of these galaxies (e.g., \cite{van95})
small 100 pc scale dust features are observed in a large
fraction of galaxies with a variety of morphologies (spiral, warped or
irregular).  \cite{pel90_} 
find that asymmetries in the isophote shapes are particularly sensitive to
the presence of dust.
Because the dust is found near the galaxy nuclei, a simple dust
screen model can underestimate the extinction or
dust column depth.  This implies that a significant fraction of light
in the nuclei could be absorbed by dust.
Despite this problem, visible WFPC and WFPC2 images from HST have been
part of a major effort to classify the nuclear stellar profiles in these
galaxies, resulting in the classification of light
profiles into two categories, galaxies with shallow inner
cusps and galaxies with `power law' light profiles
(\cite{lauer}).  We note that this classification 
is dependent on the angular resolution
of the image, and so on the distance of the galaxy.
 
Near-IR images, which are less sensitive to extinction from
dust, provide superior light profiles for dynamical
studies of early-type galaxy cores.
Although almost all early-type galaxies
harbor dust, the samples used to classify
light profiles are biased against the presence of dust precisely because
of the sensitivity of the optical images to extinction.
By using near-IR light profiles we can probe the
central light profiles for a class of ellipticals with
a higher dust content.  Since the dust itself could be related to the formation
process and subsequent evolution of the galaxy (related to merger induced 
accretion, or caused by dissipation in the gas, \cite{dubinski})
comparison of IR-observed light profiles between galaxies with varying
amounts of dust may test models of core formation.

In this paper we present the results of a survey at 1.6 microns
of early type galaxies.  We present high angular resolution
images observed with NICMOS Camera 2 on board HST.  
Surface brightness profiles and
isophotal parameters are measured from the images and
compared with the most comprehensive summary of
the visible band based studies (\cite{faber}).
We investigate the possibility of correlations between
core properties and cold gas content (traced by emission from dust
in the far infrared), hot gas content (traced by X-ray luminosity)
and cluster environment.


\section{Observations}
\subsection{The Sample}
Twenty-seven early-type galaxies, listed in Table 1, 
were observed in F160W ($1.6\mu$m) on camera 2 
of NICMOS primarily as part of a snap shot program.  
Galaxies were chosen preferentially to have existing
HST visible band images so that color maps could be made.
This included a wide variety of early type galaxies including
some galaxies with kinematically distinct cores 
(those studied in \cite{carollo}).
We then added galaxies from the RSA listed in \cite{roberts_} 
as having higher dust contents so as to insure that our sample 
was not grossly biased against dusty galaxies.
Our sample was initially chosen to 
be representative of the distribution of ellipticals
in terms of cold ISM content, however only a third
of our 80 target galaxies were observed. 
%
In Fig.~1 we show a histogram of dust masses estimated
from IRAS far infrared emission in our sample compared
to the RSA (compiled by \cite{roberts}) 
and sample of visible HST images compiled by \cite{van95_}.  
We see that our sample is not as biased
towards dust free galaxies as was the sample 
observed with WFPC and WFPC2 (compiled and discussed 
in \cite{faber}).

\subsection{The NICMOS images}
The NICMOS observation sequences were MULTIACCUM with 13 samples of 
step32 with a sequence exposure time of 192s.  
The galaxies were observed in the F160W (1.6 micron)
filter with this sequence at 4 different positions on
the sky separated by $\sim 0.56''$ resulting
in a total exposure time of 12.8 minutes per galaxy.
Images were reduced with the nicred data reduction software
(\cite{mcl})
with on-orbit flats and darks taken near the time the data
were observed.  Then each set of 4 images in a given filter
were combined according to the position observed.
The pixel size for the NICMOS camera 2 is $0.076''$.
Flux calibration for the NICMOS images
was performed using the conversion factors based on
measurements of the standard stars P330-E and P172-D during
the Servicing Mission Observatory Verification program 
(M.~Rieke, private communication).

The NICMOS images coupled with visible broad band images observed
with WFPC2 on board HST when available are shown in Fig.~2.
Galaxies lacking WFPC2 images are shown in Fig.~3.
These images show the improved ability
of the 1.6 micron images to measure stellar surface brightness
profiles in the presence of moderate amounts of dust.
For example in NGC 524, NGC 1400, NGC 1553, NGC 3056, 
NGC 4261, NGC 4278, NGC 4374, NGC 4589 and NGC 7626 
we can more accurately measure profiles from the
NICMOS images than possible with the WFPC2 images.
However we do find galaxies such as NGC 7052 and 
NGC 4150 which show evidence for large extinctions from dust
even at 1.6 microns. 
Though we can make better measurements than possible
from visible wavelength images, even in the near-IR 
the presence of the dust in these galaxies hampers
our ability to accurately measure the stellar surface brightness
profiles.

\subsection{Isophote fitting}

To measure properties of the stellar
cores, we fit ellipses to the isophotes using the ellipse routine
in the stsdas package of iraf which uses an
iterative method described by \cite{jed87}.
We also deconvolved our images with 20 iterations
of the Lucy-Richardson method (from the stsdas package of iraf) 
to take into account the point spread function of the telescope.
For deconvolution we used a model point spread function created by 
the program Tinytim 
(\cite{tinytim}) at the position of the nucleus of each galaxy on the camera.
We then reran our ellipse fitting algorithm on the deconvolved images.
The results of the ellipse fitting are shown in Fig.~4.

The ``Nuker Law'' (e.g., \cite{faber}) was then fit to 
the surface brightness profile derived from the deconvolved images.
This profile is described by
$$
I(r) = I_b 2^{(\beta-\gamma)/\alpha }
\left( {r_b \over r} \right)^\gamma
\left[1.0  +
\left( {r \over r_b} \right)^\alpha
\right]^{(\gamma -\beta)/\alpha}
$$
Here $I_b$ is the surface brightness at $r_b$
and the logarithmic slope inside the break radius
$r_b$ is $-\gamma$ and that outside is $-\beta$.
The parameters resulting from this fit are given
in Table 2.  
Surface brightness profiles are classified as either
`power law' galaxies or `core' galaxies.
To identify a galaxy as having a core we require
that the absolute value of inner logarithmic slope be 
$\gamma < 0.3$ (e.g., \cite{faber}).
We note that this classification
is dependent on the angular resolution
of the image, and so on the distance of the galaxy.
This implies that distant galaxies are more likely to
be classified as power law galaxies.

The fitted values of $r_b$ are meaningful only for the core profiles
where they represent a physical change in the nature of the
stellar distribution (see for example the discussion in \cite{faber}).
We note that our range of radii fit ($10''$ from the nucleus) 
is comparable to that of the visible band studies.  
These studies restricted the
range of their fits because outside a radius of $10''$
a deVaucouleurs law might be a better description of the
profile.
The Nuker Law was used to fit Planetary Camera images  and is intended
precisely to fit over a short range of radius.
However we note that the parameters resulting from such a fit are 
necessarily highly correlated.   
The quality of our fits is identical to that of \cite{byun_}.

\subsubsection{Comparison with previous visible band fits}

We have galaxies 14 in common with previous
HST based visible band studies (\cite{faber}, \cite{carollo}
and \cite{fer94}).
We compare the results of our fits (see Fig.~5) to those found
from these previous studies.
For the most part our break radii and exponents agree
with those found from the previous studies.  
(NGC 2636 was excluded from this comparison since no
break in surface brightness profile was listed in \cite{faber}).
Outliers (or those galaxies with fitting parameters 
that grossly disagree with those found previously) 
mostly correspond to power law galaxies where
the break radii and inner exponent value are
not particularly meaningful (NGC 1172 and NGC 1331).
%
We note that we do find a small bias towards 
measuring smaller break radii 
from the well resolved core galaxies in the NICMOS
images compared to previous visible band based studies.
We consider these cases:  NGC 1600, NGC 4374, 
NGC 4261 and NGC 5845.  

NGC 4374, NGC 4261 and NGC 5845 contain prominent
dust features within their break radii.  
The visible band images suffer
from absorption in their central regions.
This would result in measurement of a smaller
break radius or a shallower central surface brightness profile
(however the discrepancy would depend on the dust distribution).
We do not feel that the discrepancy between our measured
break radius for NGC 1600 is significant
given that the parameters of the Nuker fit are highly correlated
and that the NGC~1600 visible image used to measure surface brightness
profile properties was not a WFPC2 image and so
may be suffering from point spread function artifacts.
We find that dust features in the optical images are
the main source of discrepancies of core classification
and fitting parameters between
our work and previous visible band studies.

Figure 6 shows correlations from bulk elliptical galaxy properties
(dispersion and effective radius)
and a comparison between break radius and surface brightness
at this radius with galaxy luminosity.
As found in previous works (e.g., \cite{faber} and references therein)
the galaxies display a dichotomy:
those with measurable core break radii and low break surface brightnesses
correspond mainly to higher visual luminosity galaxies,  and those 
with steep central profiles and high central surface brightness
correspond to lower visual luminosity galaxies.
We do see a correlation between break radius and surface
brightness at this radius (as proposed in \cite{faber} and shown here
in Fig.~7).

\subsubsection{Classification between core and power law galaxies}

Our classifications (between core and power law) 
agree in all cases 
with those previous found
(listed in \cite{faber}, \cite{fer94} and \cite{carollo})
but NGC~7262, NGC~1400 and Abell 2052.

In NGC~7626 our fit to the F160W surface brightness profile
results in $\gamma=0.46$, $r_b=0''.5$.  We therefore
classify this galaxy as a power law galaxy.
However \cite{carollo_} classified it as a core galaxy 
based on WFPC2 F814W and F555W images.  
These images display 
a small warped dusty disk at $r < 0''.45$
and consequently these authors note
that their fit (with $\gamma=0$, $r_b =0''.32$)
is uncertain within $r< 0''.45$.
Since extinction from dust artificially lowers
the central surface brightness it is likely that
the NICMOS images are a better tracer of
the stellar surface brightness.  This galaxy
is likely to have the steep surface brightness profile
typical of a power law galaxy.

In NGC~1400, our fit to the F160W surface brightness
profile results in $\gamma=0.35$, $r_b=0''.85$.
This value of $\gamma$ is only slightly
above $0.3$ that divides power law from core galaxies.
\cite{faber_} classifies it as a core galaxy
(with $\gamma=0$ and $r_b =0''.33$).
However we note that the WFPC2 image does not clearly
show a core. 
In this particular case, the classification is uncertain.

Abell 2052 is quite distant and luminous 
so we expect it to be a core galaxy.  
Though we do not resolve its break radius from our images,
a shallow central profile was resolved in the WFPC2
images (\cite{byun}) which have slightly higher angular resolution
than the NICMOS images.  We therefore list it as a core galaxy.


\subsection{Ellipticity and boxiness reduction near the break radius}

In Fig.~8  we show trends observed in ellipticity and boxiness
near the break radius for the core galaxies.  
Core galaxies tend to exhibit a reduction in both boxiness (when
boxiness is exhibited) and ellipticity
between about two times the break radius and the break radius.
No galaxy is observed to be boxy within in its break radius.
Many of the cores are well resolved so this is not caused by
smoothing due to the point spread function.
There are some galaxies which stand out from this pattern, however.
NGC~7052 clearly contains a gas disk which is probably
affecting measurement of the B4 component describing boxiness
or diskiness.  NGC~1600 gains ellipticity 
and has disky (positive B4) isophotes within its core.
We suspect that this galaxy might contain a weak stellar disk.

Two processes are predicted to reduce ellipticity
and boxiness: 1) the stochasticity caused by the central black hole 
(proposed and explored by \cite{valuri}) and
2) scattering from a binary black hole.

We consider whether the morphology change in the stellar isophotes
near the break radius is consistent with the black-hole scattering
model for the formation of the cuspy or shallow core. Boxiness is a symptom 
of an uneven distribution function of
stellar orbits in phase space (\cite{binney}).
In other words this function is peaked around orbits
with a narrow range of shapes.  A scattering process would be likely to 
smooth the distribution function in phase space and so reduce boxiness.
After scattering, the angular momentum of a particle will have changed,
so we expect the distribution of scattered stars to be closer to spherical.
A population of scattered stars should therefore reduce
both boxiness and ellipticity.
A binary black hole will scatter stars at a particular
mean escape velocity (e.g.~\cite{quinlan96})
which depends on the binary semi-major axis and the binary mass ratio.
This particular velocity should then manifest as
a particular length scale over which a change in
the isophote shapes is observed.  This would naturally result
in a particular range of radius over which we see a reduction
in ellipticity and boxiness.   This would correspond 
to the region of order a few times the break radius
over which we observe the change in boxiness and
ellipticity in these two galaxies.

We now discuss an alternate possibility whereby the isophote
change is instead caused by the
the stochasticity induced by the black hole
(\cite{valuri}; \cite{merritt}).  This mechanism
reduces triaxiality (which might be consistent with the observed boxiness)
but would not likely account for such
an association with the break radius or a large change in ellipticity
as well as boxiness.
We would expect this mechanism to result in a smooth isophotal shape variation
with radius
since the diffusion timescale is primarily dependent on the local
dynamical time (\cite{ryden}).
Since we observe a shape change over a small region
(a few times the break radius),
we find that scattering from a black hole binary gives a more natural
explanation for the isophotal shape changes observed near the break radius.
However stochasticity induced by the central black hole would still
be a natural explanation for smooth shape changes observed over
larger scales (\cite{ryden}; \cite{bender+mehlert}).


\subsection{Core properties vs dust content}

The ability to observe these galaxies at 1.6 microns
allows us to study galaxies which were excluded
from visible band studies precisely because of their
dust content (e.g., NGC~4150 was excluded from \cite{faber}
precisely for this reason).
This allows us to investigate the possibility that core properties
are related to dust or cold gas content.   If dust is long lived
in elliptical galaxies then a high dust content may indicate 
a previous gas rich merger which might have resulted
in a binary black hole with more extreme binary black hole mass ratio
and a smaller core or break radius.  Alternatively if black hole coalescence
requires Gyrs to take place (\cite{begelman}) 
then we might expect recently formed
elliptical galaxies to have growing cores, again
suggesting that dusty galaxies should have smaller core or break
radii.   However, if an elliptical galaxy gains sufficient
gas then star formation can occur within its core (e.g., NGC~7052), 
a process which
might increase the central density and reduce the core size.
However as we show in Fig.~9 we see no correlation
between dust content (as measured from IRAS far-infrared emission) and 
core size.  This suggests 
that the dust or cold gas we now see in elliptical
galaxy is acquired subsequent to core formation and does not
strongly affect the surface brightness profile.
This is consistent with the lack of correlation between
dust quantity and galaxy properties such as luminosity
(\cite{goodfrooij}), and the lack of alignment of dust features 
with respect to galaxy axes which suggests  that dust must be acquired in
elliptical galaxies on relatively short timescales (\cite{van95}).    

Since we see no
strong color gradients in our color maps we find no evidence for 
a diffuse dust component that is distributed 
in a different way than the light density.
If such a component exists it must be extremely diffuse and
similarly distributed as the star light (\cite{goodfrooij}).


\subsection{Core properties vs X-ray luminosity}

We also investigate the possibility that X-ray luminosity
may be related to elliptical core properties
(see Fig.~10).
X-ray and optical luminosities of early-type galaxies
are correlated with $L_X \propto L_B^{2.0\pm 0.2}$ (\cite{eskridge_3}).
However the scatter about this relation is enormous
with $L_X/L_B$ varying by factors of 500.
There are two approaches towards accounting for this large
scatter: 1) evolution models for production of 
hot gas via supernovae (e.g., \cite{ciotti})
and 2) consideration of environmental affects such
as ram pressure stripping (e.g., {white}).
In either scenario the formation
and evolution of the core might be related, so we
might expect a correlation between core properties
and X-ray luminosity.

However, we fail to see any strong correlation between core properties 
and X-ray luminosity or $L_X/L_B$ (see Fig.~7).
In our sample we find a core galaxy in a poor group, NGC~524,
which moderate X-ray flux as well as X-ray bright cluster
ellipticals.  
The lack of correlation 
suggests that the process of core formation is subsequently
unaffected by whatever processes determine the X-ray luminosities
of galaxies.

\subsection{Core properties vs environment}

Our sample combined with that of \cite{faber_} contain galaxies
which span a range of environments from poor groups
to moderately sized clusters such as the Virgo and Fornax 
Clusters.  We find poor groups with brightest
members with cores (NGC~524)
and poor groups with brightest members without cores
(NGC~1553, NGC~2907, NGC~5198, NGC~821, NGC~1172, NGC~1400, NGC~7626).  
(We have identified cluster membership based on \cite{garcia}, \cite{faber89}
and references therein.)
Almost all the brighter galaxies ($M_V < -20.5$) 
in Virgo and Fornax contain cores (see the histograms
presented in Fig.~11).
However galaxies of this luminosity or greater without cores 
appear to be common in poorer environments
(e.g., NGC~821, NGC~1172, NGC~1400, NGC~1553, NGC~1700, NGC~3115, 
NGC~4594, NGC~4697, NGC~5198).
As we see in Table 3 and Fig.~7 these are not primarily 
distant galaxies which would have cores
that are unresolved (except possibly in the case of NGC~5198). 
Since the scatter from the fundamental plane for these galaxies
(\cite{faber})
is much smaller than a magnitude, distance errors are unlikely
to account for the difference in the histograms between
the Virgo and Fornax cluster galaxies and the whole
sample (Figs.~11a and 11b).
This suggests that poor galaxy groups can harbor more luminous 
power law galaxies than clusters.  
This is an interesting possibility that should be investigated
further with bigger samples. 

\section{Summary and Discussion}

In this paper we have presented a NICMOS imaging
study of early-type galaxies.  In moderately dusty
galaxies these
images allow us to measure stellar surface brightness
profiles more accurately than possible with visible band images
which are more strongly affected by extinction from
dust.  
Discrepancies between core classification 
and measurement of core properties between
our NICMOS images and previous visible band based HST
studies are primarily due to dust features in the nuclear
regions of the galaxies.

We observe a trend in boxiness and ellipticity
in the core galaxies.  Both boxiness and ellipticity
are reduced near and within the break radius.
No galaxy is observed to be boxy within its break radius.
This is consistent with a core formation mechanism that
involves scattering of stars (such as scattering from a binary
black hole).

We failed to find correlations between 
core break radius and dust content or X-ray luminosity.  
This suggests that the current cold or hot gas content 
of an elliptical galaxy is unrelated to the process
of core formation. The gas content is then more likely to be determined
by processes, such as minor mergers and cooling flows,
that would occur after the formation of the galaxy or core.

By combining our sample with that of \cite{faber_} 
we find that galaxies from the Virgo and Fornax clusters
(together) show a dichotomy of core types which is strongly dependent
on luminosity.  However,  the dependence of core type
on luminosity may be weaker in the complete sample including
galaxies outside of the Virgo and Fornax Clusters.
In particular higher luminosity power law galaxies
may be more common in poorer environments.
Since both core classification and cluster identification
tequniques are strongly dependent
on the galaxy distance, care must be taken to ensure
that the more luminous power law galaxies are not the most distant.
A larger sample of galaxies 
(such as is now available in the HST archive)
could more thoroughly
probe the relation between core properties and environment.

\acknowledgments

Support for this work was provided by NASA through grant number
GO-07886.01-96A
from the Space Telescope Institute, which is operated by the Association
of Universities for Research in Astronomy, Incorporated, under NASA
contract NAS5-26555.
We also acknowledge support from NASA project NAG-53359.
We acknowledge helpful discussions and correspondence with
A.~Alonso-Herrero, M.~Elvis, M.~Rieke, G.~Rieke,
S.~Stolovy, N.~Caldwell and J.~Bechtold.
We thank M.~S.~Roberts for providing catalogues to us in electronic form.


\clearpage


\vfill\eject



\def\smm{\mathrel{\setminus}}

 
\begin{deluxetable}{lcccccccc}
\footnotesize
\tablewidth{460pt}
\tablecaption{Sample of Early-Type Galaxies }
\tablehead{
\colhead{Galaxy} & 
\colhead{Type} & 
\colhead{B$_T^0$} & 
\colhead{V$_{\rm sys}$} & 
\colhead{D} & 
\colhead{$\sigma$}  \\
\colhead{} & 
\colhead{} & 
\colhead{mag} & 
\colhead{(km s$^{-1}$)} & 
\colhead{(Mpc)} & 
\colhead{(km s$^{-1}$)} \\
\colhead{(1)} & 
\colhead{(2)} & 
\colhead{(3)} & 
\colhead{(4)} & 
\colhead{(5)} & 
\colhead{(6)} 
}
\startdata
    NGC524 &       S0     & 11.3 &  2416 &  23.1 &  270 \nl
    NGC821 &           E6 & 11.9 &  1716 &  28.6 &  215 \nl
   NGC1052 &        E3/S0 & 11.5 &  1475 &  21.5 &  204 \nl
   NGC1172 &     S0       & 12.6 &  1669 &  29.8 &  113 \nl
   NGC1331 &           E2 & 14.1 &  1375 &  22.0 &      \nl
   NGC1351 &    S0        & 12.6 &  1527 &  17.8 &  144 \nl
   NGC1400 &       S0     & 11.6 &   549 &  21.5 &  269 \nl
   NGC1426 &           E4 & 12.2 &  1443 &  21.5 &  157 \nl
   NGC1553 &  S0          & 10.4 &  1280 &  16.0 &  184 \nl
   NGC1600 &           E4 & 11.8 &  4687 &  50.2 &  323 \nl
   NGC2636 &           E0 & 14.6 &  1896 &  33.5 &   85 \nl
   NGC2907 &    S0        & 12.8 &  2090 &  26.0 &      \nl
   NGC3056 &     S0       & 12.6 &  1017 &  13.0 &      \nl
   NGC4150 &    S0        & 12.4 &   244 &   4.0 &   85 \nl
   NGC4261 &           E3 & 11.4 &  2200 &  35.0 &  339 \nl
   NGC4278 &           E1 & 11.1 &   643 &  18.4 &  243 \nl
   NGC4291 &           E3 & 12.3 &  1715 &  37.9 &  295 \nl
   NGC4374 &           E1 & 10.2 &  1033 &  15.3 &  296 \nl
   NGC4589 &           E2 & 11.8 &  1985 &  37.9 &  241 \nl
   NGC4636 &    E0/S0     & 10.2 &   937 &  15.3 &  217 \nl
   NGC5198 &           E1 & 12.9 &  2569 &  46.6 &  212 \nl
   NGC5845 &            E & 13.4 &  1450 &  28.2 &  244 \nl
   NGC5982 &           E3 & 12.0 &  2936 &  39.0 &  248 \nl
   NGC7052 &           E4 & 12.0 &  4920 &  61.5 &  275 \nl
   NGC7626 &           E1 & 12.2 &  3416 &  44.8 &  270 \nl
ESO507-G045 &           S0 & 12.0 &  4825 &  60.3 &  325 \nl
     A2052 &            E & 13.7 & 10200 & 127.5 &  238 \nl
\enddata
\tablecomments{
Columns: (1) Galaxy. 
(2) Hubble Type.
(3) B band magnitude corrected for Galactic extinction
for the galaxy (or bulge for the S0s)
from Faber et al.~(1987) or from the RC3.
(4) Heliocentric systemic velocity.
(5) Adopted distance taken from Faber et al.~(1997) and references
therein or from the 
heliocentric velocity assuming $H_0 = 80$ km s$^{-1}$ Mpc$^{-1}$. 
(6) Velocity dispersions taken from Faber et al.~(1989)
or McElroy (1995).
}
\end{deluxetable}

\clearpage

\begin{deluxetable}{lccccccccccc}
\footnotesize
\tablewidth{460pt}
\tablecaption{Nuker Fits and Major Axis PA }
\tablehead{
\multicolumn{1}{l}{Galaxy} &
\colhead{Core Type } & 
\colhead{$I_b^{lim}$}  & 
\colhead{$I_b$}  & 
\colhead{$\beta$}   & 
\colhead{$\gamma$}  & 
\colhead{$\alpha$}  &  
\colhead{$r_b$}     &
\colhead{ORIENTAT}  &
\colhead{PA Major Axis} \\
\colhead{}          & 
\colhead{}          & 
\multicolumn{2}{c}{mag/arcsec$^2$}  & 
\colhead{}          & 
\colhead{}          & 
\colhead{}          & 
\colhead{arcsec}    &
\colhead{degree}  &
\colhead{degree}    \\
\colhead{(1)}        & 
\colhead{(2)}        & 
\colhead{(3)}        & 
\colhead{(4)}        & 
\colhead{(5)}        & 
\colhead{(6)}        & 
\colhead{(7)}        & 
\colhead{(8)}        & 
\colhead{(9)}        & 
\colhead{(10)}         
}
\startdata
%
   NGC0524 & $\cap$ &      & 13.29 &   1.34 &   0.25 &   0.93 &   1.10 &   21.67 &  165.57 \nl 
   NGC0821 & $\smm$ &10.72 & 13.14 &   1.47 &   0.78 &   1.20 &   1.25 &   31.18 &  149.53 \nl 
   NGC1052 & $\cap$ &      & 11.77 &   1.27 &   0.18 &   2.16 &   0.44 &   56.81 &  143.89 \nl 
   NGC1172 & $\smm$ &11.35 & 12.64 &   1.49 &   0.62 &   0.78 &   0.32 &   40.94 &  137.09 \nl 
   NGC1331 & $\smm$ &13.76 & 16.47 &   1.29 &   0.57 &   3.44 &   2.91 &   89.30 &   16.40 \nl 
   NGC1351 & $\smm$ &11.42 & 13.41 &   1.31 &   0.78 &   3.78 &   1.00 &  110.28 &   99.92 \nl 
   NGC1400 & $\smm$?&11.35 & 12.74 &   1.52 &   0.35 &   1.06 &   0.86 &   57.53 &  100.87 \nl 
   NGC1426 & $\smm$ &11.18 & 13.55 &   1.28 &   0.84 &   2.51 &   1.10 &   80.69 &  123.61 \nl 
   NGC1553 & $\smm$ &10.77 & 14.08 &   1.43 &   0.74 &   3.82 &   5.72 &   59.74 &   30.26 \nl 
   NGC1600 & $\cap$ &      & 14.46 &   0.91 &   0.04 &   2.50 &   1.61 &   90.91 &    8.99 \nl 
   NGC2636 & $\smm$ &12.57 & 16.85 &   1.72 &   1.03 &   1.26 &   3.61 &  133.51 &   46.49 \nl 
   NGC2907 & $\smm$ &10.99 & 13.75 &   1.78 &   0.58 &   0.52 &   1.96 &   54.95 &    3.75 \nl 
   NGC3056 & $\smm$ &11.28 & 15.20 &   1.80 &   0.90 &   2.07 &   4.11 &  -72.14 &  165.14 \nl 
   NGC4150 & $\smm$ &11.05 & 11.28 &   1.17 &   0.82 &   8.96 &   0.13 &   99.17 &   80.83 \nl 
   NGC4261 & $\cap$ &      & 13.51 &   1.39 &   0.16 &   2.47 &   1.85 &  -94.05 &  165.85 \nl 
   NGC4278 & $\cap$ &      & 12.70 &   1.40 &   0.03 &   1.70 &   1.12 &   96.29 &    4.01 \nl 
   NGC4291 & $\cap$ &      & 12.01 &   1.40 &   0.11 &   2.21 &   0.44 & -169.46 &   84.16 \nl 
   NGC4374 & $\cap$ &      & 13.17 &   1.49 &   0.15 &   2.12 &   2.45 & -116.33 &    3.03 \nl 
   NGC4589 & $\cap$ &      & 12.03 &   1.19 &   0.09 &   1.51 &   0.28 & -179.16 &  102.36 \nl 
   NGC4636 & $\cap$ &      & 14.20 &   1.33 &   0.16 &   2.06 &   2.85 & -114.18 &  114.18 \nl 
   NGC5198 & $\smm$ &11.69 & 12.01 &   1.11 &   0.88 &   1.29 &   0.17 & -136.71 &   81.71 \nl 
   NGC5845 & $\smm$ &10.72 & 13.01 &   2.18 &   0.51 &   2.11 &   1.38 & -132.42 &   48.12 \nl 
   NGC5982 & $\cap$ &      & 13.01 &   1.43 &   0.19 &   1.07 &   0.80 & -162.04 &   72.04 \nl 
   NGC7052 & $\cap$ &      & 13.50 &   1.19 &   0.16 &   2.76 &   0.77 &  -62.87 &   11.27 \nl 
   NGC7626 & $\smm$ &11.94 & 12.86 &   1.15 &   0.47 &   2.79 &   0.51 &    6.38 &  162.12 \nl 
   ESO507  & $\cap$ &      & 12.49 &   1.26 &   0.16 &   1.34 &   0.33 &  -95.79 &  176.99 \nl 
   A2052   & $\cap$?&14.26 & 15.08 &   0.49 &   0.54 &   1.00 &   0.43 & -126.74 &  110.64 \nl 
\enddata
\tablecomments{
Columns: (1) Galaxy. 
(2) Core classification.  $\cap$ refers to galaxies
with core type surface brightness profiles.  $\smm$ refers to galaxies with 
power law profiles. 
(3) $I_b$ is the surface brightness at the break radius, $r_b$,
given in H mag/arcsec$^2$ (calibration of the F160W images is from M.~Rieke
private communication).  
(4) For power law galaxies only our angular resolution sets
an upper limit on a radius inside which a shallow core could exist
$r_b^{lim} < 0.1''$. We denote
$I_b^{lim}$ as a limit on the surface brightness 
at this radius.
(5-7) Exponents resulting from a fit to the `Nuker' surface
brightness profile law.  
(8) Core break radius. 
(9) Position angle of F160W image y axis (degrees east of north).
(10) Mean major axis position angle (degrees east of north).
}
\end{deluxetable}

\begin{deluxetable}{lccccccccccc}
\footnotesize
\tablewidth{460pt}
\tablecaption{Derived Quantities}
\tablehead{
\multicolumn{1}{l}{Galaxy} &
\colhead{$M_V$ }         &
\colhead{log $r_b$}      &
\colhead{log $r_b^{lim}$}   &
\colhead{log $M_{dust}$} &
\colhead{log $L_B$}      &
\colhead{log $L_X$}      \\
\colhead{}                    &
\colhead{mag}                 &
\colhead{log(pc)}             &
\colhead{log(pc)}             &
\colhead{log($M_\odot$)}      &
\colhead{log(ergs/s)}         &
\colhead{log(ergs/s)}         \\
\colhead{(1)}            &
\colhead{(2)}            &
\colhead{(3)}            &
\colhead{(4)}            &
\colhead{(5)}            &
\colhead{(6)}            &
\colhead{(7)}            
}
\startdata
%
    NGC524 & -21.4 &  2.09 &    & 5.45 &  43.77 &  40.43 \nl
    NGC821 & -21.3 &  2.24 &1.14& 5.97 &  43.73 &        \nl
   NGC1052 & -21.0 &  1.66 &    & 5.00 &  43.63 &  40.49 \nl
   NGC1172 & -20.7 &  1.66 &1.16&      &  43.50 &  39.92 \nl
   NGC1331 & -18.4 &  2.49 &1.03&      &  42.60 &        \nl
   NGC1351 & -19.4 &  1.94 &0.94& 5.26 &  43.01 &        \nl
   NGC1400 & -21.1 &  1.95 &1.02& 5.99 &  43.59 &  40.03 \nl
   NGC1426 & -20.3 &  2.06 &1.02&      &  43.34 &        \nl
   NGC1553 & -21.5 &  2.65 &0.89& 4.69 &  43.84 &  40.42 \nl
   NGC1600 & -22.7 &  2.59 &    & 4.88 &  44.26 &  41.36 \nl
   NGC2636 & -18.9 &  2.77 &1.21&      &  42.76 &        \nl
   NGC2907 & -20.2 &  2.39 &1.10& 5.65 &  43.27 &        \nl
   NGC3056 & -18.8 &  2.41 &0.80&      &  42.77 &        \nl
   NGC4150 & -16.3 &  0.40 &0.29& 3.90 &  41.82 &        \nl
   NGC4261 & -22.3 &  2.50 &    & 4.42 &  44.11 &  41.13 \nl
   NGC4278 & -21.2 &  2.00 &    & 5.33 &  43.65 &        \nl
   NGC4291 & -21.6 &  1.90 &    &      &  43.82 &  41.05 \nl
   NGC4374 & -21.6 &  2.26 &    & 4.74 &  43.85 &  40.64 \nl
   NGC4589 & -22.0 &  1.71 &    & 5.53 &  44.01 &  40.52 \nl
   NGC4636 & -21.7 &  2.33 &    & 4.57 &  43.86 &  41.25 \nl
   NGC5198 & -21.4 &  1.58 &1.36&      &  43.75 &        \nl
   NGC5845 & -19.9 &  2.28 &1.14&      &  43.13 &        \nl
   NGC5982 & -21.8 &  2.18 &    & 6.12 &  43.94 &  40.68 \nl
   NGC7052 & -22.9 &  2.36 &    & 6.48 &  44.35 &        \nl
   NGC7626 & -22.0 &  2.05 &1.34&      &  44.01 &  40.98 \nl
    ESO507 & -22.8 &  1.98 &    &      &  44.33 &        \nl
     A2052 & -22.7 &  2.43 &1.79&      &  44.30 &        \nl
\enddata
\tablecomments{
Columns: (1) Galaxy.
(2) Absolute visual magnitude.
(3) The log of the core break radius in pc from the Nuker fits 
listed in Table 2.
(4) For the core galaxies, the log of the maximum break radius (in pc) 
for a shallow core based on an angular resolution of $0.1''$.
(5) Dust mass derived from IRAS fluxes using the flux densities
and relation given in Roberts et al.~(1991) and the distances
listed in Table 1.
(6) The B band luminosity ($\nu F_{\nu}$ at $0.44\mu$m)
estimated from $M_V$ using $B-V$ colors from 
Faber et al.~(1989) or from the RC3.
(7) The X-ray  luminosity in the energy range 0.5-4.5 kev
from the Einstein IPC or HRI 
observations estimated from fluxes tabulated in Roberts et al.~(1991).
}
\end{deluxetable}

\clearpage
\end{document}